\documentclass[aps,twocolumn,showpacs,showkeys]{revtex4}
\usepackage{graphicx}
\usepackage{amssymb}
\usepackage{bm}

\begin{document}
\setcounter{page}{1}
\title{Percolation Transition in Correlated Static Model}

\author{Sang-Woo \surname{Kim}}
\affiliation{Department of Physics, University of Seoul,
  Seoul 130-743, Korea}
\author{Jae Dong \surname{Noh}}
\email{jdnoh@uos.ac.kr}
\thanks{Fax: +82-2-2245-6531}
\affiliation{Department of Physics, University of Seoul,
  Seoul 130-743, Korea}

\date[]{Received 13 September 2007}
\begin{abstract}
We introduce a correlated static model and investigate a percolation 
transition. The model is a modification of the static model and is 
characterized by assortative degree-degree correlation. 
As one varies the edge density,
the network undergoes a percolation transition. The percolation transition
is characterized by a weak singular behavior of the mean cluster size and
power-law scalings of the percolation order parameter and the cluster size
distribution in the entire non-percolating phase. These results
suggest that the assortative degree-degree correlation generates a global
structural correlation which is relevant to the
percolation critical phenomena of complex networks.
\end{abstract}

\pacs{89.75.Hc, 05.10.-a, 05.70.Fh, 05.50.+q}
\keywords{Percolation, Degree-degree correlation, Power law, Critical phenomena}

\maketitle

\section{Introduction}\label{sec:1}
Many systems in nature have a complex network structure. For example, the
Internet is a wired network of computers and routers, the World Wide Web
is a network of web pages hyperlinked to each other, a protein interaction
network is a network of interacting proteins in an organism, and 
a social network is a network of individuals who are linked through 
a certain relationship.  Such complex networks have a highly inhomogeneous
structure. In order to characterize and understand the structure and
dynamics, extensive research has been performed for the last
decade~\cite{Watts98,Albert02,Dorogovtsev02,Newman03,SHLee06,Noh06,Noh07c}.

A degree distribution $p(k)$ for the probability of a node having degree
$k$ is a quantity of primary importance. It is one of the
essential characteristics that influences structural properties, dynamical
behaviors, and collective phenomena of complex networks.
However, the degree is a property of each individual vertex. 
Hence, the degree
distribution by itself cannot describe correlation among different vertices.
It is attracting growing interest, since many real-world networks
display a certain level of correlation, which has a significant impact on
network properties~\cite{Newman02,Vazquez03,Serrano06,Rozenfeld07}.

The degree-degree~(DD) correlation refers to the correlation 
between degrees of neighboring vertices~\cite{Newman02}. 
The whole information on the correlation is contained in the degree correlation 
function $p(k',k)$, the probability of an edge linking nodes of degree 
$k$ and $k'$, and the  conditional probability $p(k'|k) \equiv p(k',k) /
(\sum_{k''} p(k'',k))$, the probability of a
node among neighbors of degree-$k$ nodes having degree $k'$.
The overall feature is conveniently characterized by the assortativity and
the mean neighbor degree function. 
The assortativity coefficient $r$ is defined as the normalized Pearson 
correlation coefficient between degrees of neighboring 
vertices~\cite{Newman02}, and the mean neighbor degree function is defined
as $K_{NN}(k) \equiv \sum_{k'} k' p(k'|k)$~\cite{Pastor-Satorras01,Vazquez02}.
A network with a positive~(negative) value of $r$ or an
increasing~(decreasing) function $K_{NN}(k)$ 
has a positive~(negative) correlation and is called 
assortative~(disassortative). A network with $r=0$ or a constant function
$K_{NN}(k)$ is called uncorrelated or neutral.

Recent studies have shown that the structural correlation, as well as the
degree distribution, plays an important role in percolation critical
phenomena~\cite{Rozenfeld07,Noh07a}. 
In Ref.~\cite{Noh07a}, the percolation transition was studied in the
exponential random graph~(ERG) model.
With the ERG model one can simulate
assortative, neutral, or disassortative networks with the same degree
distribution. 
The numerical study reveals that the disassortative and the neutral ERG models 
display the percolation transition in the same universality class.
On the other hand, the assortative ERG model displays the percolation 
transition in a distinct universality class. The 
percolation transition in the assortative ERG model is similar in nature 
to that in the growing-network 
models~\cite{Callaway01,Dorogovtsev01,JKim02,Krapivsky04}
which are also assortative.
It is noteworthy that the percolation transition in the assortative 
ERG model cannot be explained only with the DD
correlation~\cite{Noh07b}. 
A possible scenario is that the assortative DD
correlation may give rise to a global structural correlation that is relevant
to the percolation critical phenomena.

In this work, we introduce a model for assortative networks 
and investigate the percolation transition of the model. 
The purpose is to confirm whether the assortative DD correlation is relevant
to the percolation critical phenomena. Another purpose is to present an
efficient stochastic model of correlated networks with degree
distribution ranging from the Poisson distribution to scale-free power-law
distributions.
The ERG model introduced in Ref.~\cite{Noh07a} is based on a Monte Carlo
method which takes a longer time to generate large-size networks.
We present a model with which one can generate a correlated network 
fast and easily. This model can also be used for further studies of critical
phenomena other than the percolation.

In Sec.~\ref{sec:2}, we introduce a model for networks with an assortative DD
correlation. This is a modification of the so-called static
model~\cite{Goh01} which is a model for uncorrelated scale-free networks. 
Our model will be referred to as the correlated static model. 
Basic properties of the model are also presented.
In Sec.~\ref{sec:3}, we investigate bond percolation transitions in the
correlated static model. We summarize and conclude the paper in 
Sec.~\ref{sec:4}. 

\section{Correlated static model}\label{sec:2}
The static model is an efficient model for uncorrelated
networks~\cite{Goh01}. 
A static-model network with $N$ vertices and $K$ edges is
constructed as follows: (i) Each vertex $i$~($i=1,\cdots,N)$ 
is assigned to a weight $w_i = i^{-\mu}/(\sum_{j=1}^N j^{-\mu})$; 
(ii) according to the probability $\{w_i\}$, two vertices are chosen at
random.
If there is no edge between them, they are linked with an edge.
The procedure (ii) is repeated until one has $K$ edges in total. A resulting
network is scale-free with a power-law degree distribution $p(k) \sim
k^{-\lambda}$ with the degree exponent $\lambda=1+1/\mu$. When $\mu=0$, all
vertices are chosen with equal probability and the model 
reduces to the Erd\H{o}s-R\'enyi random network with the Poisson degree 
distribution. The static-model network is uncorrelated for
$\mu<1/2$ or $\lambda>3$~\cite{Lee06}. The percolation transition in this
model has been thoroughly studied~\cite{Lee04}.

We modify the static model to incorporate the assortative DD
correlation. A network with $N$ vertices and $K$ edges
in the correlated static model is constructed as follows:
(i) Each vertex $i$~($i=1,\cdots,N$) is assigned to a weight
$w_i=i^{-\mu}/(\sum_{j=1}^N j^{-\mu})$; (ii) a vertex $i$ is chosen with 
probability $w_{i}$ at random, and
then another vertex $j$ is chosen randomly among all
vertices with the same degree as $i$. If there is no edge between $i$ and
$j$, an edge connecting them is added. The procedure (ii) is repeated
until there are $K$ edges in total. 
Note that the static model and the correlated
static model differ in the procedure (ii). While edges are added between 
vertices chosen independently in the former, they are added between 
vertices of the same degree in the latter. This generates a positive DD
correlation.

Consider the degree distribution of the correlated static model. Let
$p_i(k,t)$ be the probability that a vertex $i$ has degree $k$ at time step 
$t$. Since an edge is added at each time step, the total number of edges $K$
is equal to $t$.
The degree distribution is given by $p(k,t) = \frac{1}{N}\sum_{i}p_i(k,t)$.
For the sake of convenience, we introduce $q_k(t) = \sum_i w_i p_i(k,t)$.
Then, in the large-$N$ limit, the time evolution of $p_i(k,t)$ is governed by
\begin{eqnarray*}
p_i(k,t+1) - p_i(k,t) &=& \left[ w_i +\frac{q_{k-1}(t)}{Np_{k-1}(t)} 
               \right] p_i(k-1,t) \\
           && - \left[ w_i +\frac{q_{k}(t)}{Np(k,t)} \right] p_i(k,t) ,
\end{eqnarray*}
with the initial condition $p_i(k,t=0) = \delta(k,0)$.
The terms in the square brackets become $2w_i$ for the static model.
Summing up both sides over all $i$, one finds the evolution equation 
of the degree distribution function $p(k,t)$. This is given by
\begin{equation}
p(k,t+1) = p(k,t) + \frac{2}{N} \left[ q_{k-1}(t) - q_k(t)\right].
\end{equation}
Note that the correlated static model has the same evolution equation 
as the static model. Hence we conclude that the correlated static model 
has the same degree distribution as the static model. From the properties of
the static model~\cite{Goh01}, follows the
power-law degree distribution 
\begin{equation}
p(k) \sim k^{-\lambda}
\end{equation} 
with the degree exponent 
\begin{equation}
\lambda=1+1/\mu
\end{equation}
for $\mu>0$ and the Poisson distribution at $\mu=0$.

\begin{figure}[t!]
\includegraphics*[width=\columnwidth]{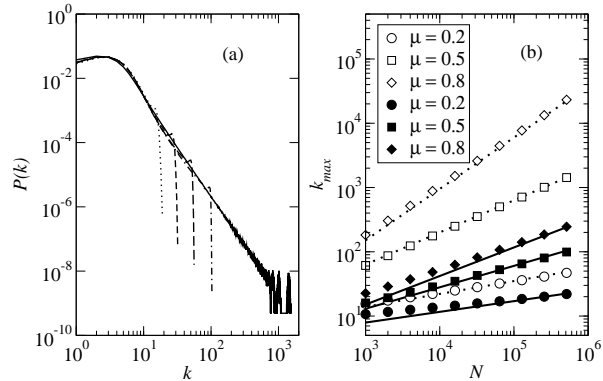}
\caption{(a) Degree distributions of the static model~(solid line) 
with $N=2^9 \times 10^3$ and the correlated static 
model with $N=10^3$~($\cdots$), $2^3 \times 10^3$~(- - -),
$2^6\times 10^3$~(-- -- --), and $2^9\times 10^3$~(-$\cdot$-$\cdot$-) 
with $\mu=0.5$ and link density $\delta=K/N=2.0$.
(b) Scaling of the maximum degree at $\mu=0.2$, $0.5$, and $0.8$. 
Open symbols are for the static model and closed symbols are for 
the correlated static model.
Solid lines have slope $1/\lambda=\mu/(1+\mu)$ and dotted lines have slope
$1/(\lambda-1)=\mu$.}
\label{fig1}
\end{figure}

We present the degree distributions of the correlated static model and the
static model with $\mu=0.5$ in Figure~\ref{fig1}(a). 
This confirms the theoretical prediction that both models have the power-law
degree distribution with the same degree exponent $\lambda$.
However, we find that the maximum cutoff degree $k_{max}$
scales differently with the network size $N$.
The maximum degree $k_{max}$ is known to scale as 
$k_{max} \sim N^{1/(\lambda-1)}$ in the static model~\cite{Lee04}.
This can be derived from the condition that the number of nodes with
$k>k_{max}$ should be of the order of unity. This condition demands
that $\int_{k_{max}}^\infty p(k)dk \sim 
\int_{k_{max}}^\infty k^{-\lambda} dk = \mathcal{O}(1/N)$, which leads to
$k_{max} \sim N^{1/(\lambda-1)}$.
On the other hand, as one can see in Figure~\ref{fig1}(b), $k_{max}$ in the
correlated static model is much smaller than that in the static model. 
The following argument explains the scaling behavior of $k_{max}$.
In the correlated static model, one can add a link to a vertex 
only if there exists another node with the same degree. So, if a node has
the maximum degree $k_{max}$, there should be another node having the
same degree $k_{max}$. 
This leads to the constraint $p(k_{\max}) = \mathcal{O}(1/N)$ 
instead of $\int_{k_{max}}^\infty p(k) dk =
\mathcal{O}(1/N)$.  This gives
\begin{equation}\label{k_max}
k_{max} \sim N^{1/\lambda} \ .
\end{equation}
The numerical data in Figure~\ref{fig1}(b) are in good agreement
with the scaling behavior in Eq.~(\ref{k_max}).

\begin{figure}[t!]
\includegraphics*[width=\columnwidth]{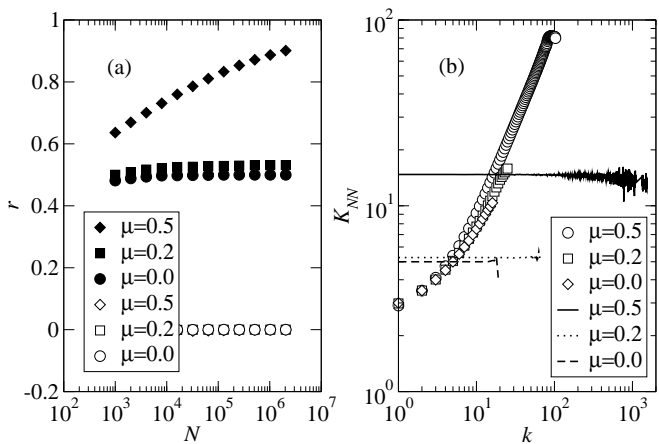}
\caption{(a) Assortativity coefficient of the static model~(open symbols)
and the correlated static model~(closed symbols). (b) $K_{NN}$ {\it vs.} 
$k$ in the
static model~(lines) and the correlated static model~(symbols) with
$N=512000$ vertices.}\label{fig2}
\end{figure}

In order to examine the DD correlation, we measure the assortativity
coefficient $r$~\cite{Newman02}. This is defined as
\begin{equation}
r = \frac{ \langle k_{e,1} k_{e,2} \rangle_e - \langle
(k_{e,1}+k_{e,2})/2\rangle_e^2}{ \langle (k_{e,1}^2 + k_{e,2}^2)/2 \rangle_e
- \langle (k_{e,1}+k_{e,2})/2\rangle_e^2 } ,
\end{equation}
where $\langle (\cdot)\rangle_e$ represents the average over all edges, and
$k_{e,1}$ and $k_{e,2}$ denote the degrees of two vertices connected with an
edge $e$. It is measured and presented in Figure~\ref{fig2}(a). 
While the assortativity coefficients vanish as $N$ increases in the static
model, they converge to a finite value as $N$ increases in the correlated
static model. Positive correlation is also observed in the mean neighbor
degree $K_{NN}(k)$ presented in Figure~\ref{fig2}(b). In the correlated
static model $K_{NN}(k)\sim k$, while $K_{NN}(k)$ is a constant function in
the static model. This shows that the correlated static model has the desired
property. 
\begin{figure}[t]
\includegraphics*[width=\columnwidth]{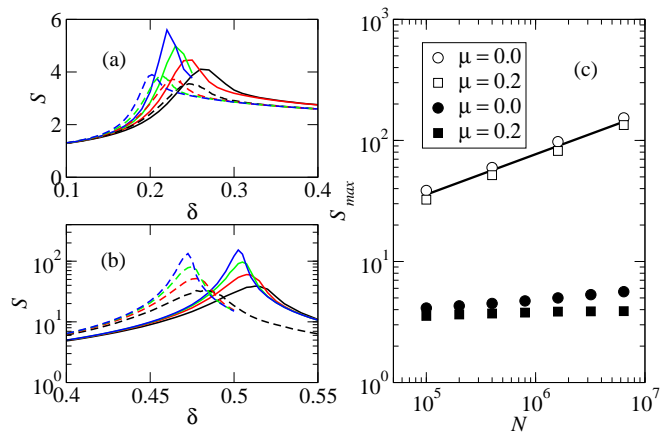}
\caption{(a) $S$ {\it vs.} $\delta$ for the correlated static model (a) and 
the static model (b) at $\mu=0$~(solid lines) and $0.2$~(dotted lines). 
Network sizes are $N=10^5$, $4\times 10^5$, $16\times 10^5$, and
$64\times 10^5$. The larger $N$ is, the higher the peak is.
(c) Peak height $S_{max}$ {\it vs.} $N$. Closed
symbols are for the correlated static model, and open symbols are for the
static model. The solid line has slope $1/3$.}
\label{fig3}
\end{figure}

\section{Percolation transition}\label{sec:3}
As a static-model network does~\cite{Lee04}, so also
a correlated model network undergoes a
percolation transition as one increases the edge density $\delta\equiv K/N$.
To a given value of $\delta$, 
vertices are decomposed into disjoint sets~(called clusters)
in such a way that all vertices in a cluster are mutually connected, while 
those in different clusters are not. The size $s$ of a cluster is defined 
as the number of vertices in it. 
When there is no edge~($\delta=0$), each vertex belongs to a cluster of size 
$s=1$. As edges are added, small clusters merge into large ones. A cluster
configuration can be characterized by the cluster size distribution $n_s$, 
which is defined as the ratio of the number of clusters of size $s$ to $N$.

The order parameter for the percolation transition is given by
\begin{equation}
P_\infty \equiv \frac{s_{max}}{N},
\end{equation}
where $s_{max}$ is the size of the largest cluster.
A network is said to be in a {\em percolating} phase if $P_\infty$ is finite 
in the
$N\rightarrow \infty$ limit. In that case, the largest cluster is called
the infinite or giant cluster. If $P_\infty$ vanishes in the
$N\rightarrow \infty$ limit, a network is said to be in a {\em
non-percolating} phase. 
Another useful quantity is the mean cluster size $S$, given by
\begin{equation}
S \equiv \frac{ \sum'_s s^2 n_s }{\sum'_s s n_s} ,
\end{equation}
where the summation is over all clusters except for the largest cluster.

In the static model, the percolation transition
is characterized by power-law scalings, $P_\infty \sim
(\delta-\delta_c)^\beta$ and $S\sim|\delta-\delta_c|^{-\gamma}$ for
$\lambda>3$, where $\delta_c = (1-2\mu)/(2(1-\mu)^2)$ is the 
percolation threshold. The critical
exponents have the values $\beta=1$ and $\gamma=1$ for $\lambda>4$, and
$\beta=1/(\lambda-3)$ and $\gamma=1$ for $3<\lambda<4$~\cite{Lee04}.
When $\lambda\leq 3$, the networks are always in the percolating phase.

We perform numerical studies of the percolation transition in the
correlated static model and compare the results with those of the static
model. A striking difference is observed in the behavior of the mean
cluster size $S$. In Figure~\ref{fig3}, we compare 
$S$ obtained for the static and correlated static models 
at $\mu=0$ and $0.2$. 
In the static-model cases, there are peaks in the plot of $S$ near
percolation thresholds, which sharpen as $N$ increases. 
Theoretically, the peak heights should grow
algebraically as $S_{max} \sim N^{\gamma/\bar{\nu}}$ with the
finite-size-scaling exponent $\bar{\nu}=3$ for $\lambda>4$ and
$\bar{\nu}=(\lambda-1)/(\lambda-3)$ for $3<\lambda<4$~\cite{Lee04}. 
Our data are consistent
with the power-law scaling for the static model. 

On the other hand, in the correlated-static-model cases, the peaks are not 
as sharp as in the static model. 
The peak heights seem to converge to a finite value or 
at most scale logarithmically with $N$. 
The weak singular behavior of $S$ suggests 
that the percolation transition in the
correlated static model does not belong to the same universality as that in
the uncorrelated static model.

In order to characterize the phase transition, we study 
finite-size-scaling behaviors of the order parameter $P_\infty$ near
percolation thresholds. At 
criticality, we expect that $P_\infty$ scales algebraically as
\begin{equation}
P_\infty (N) \sim N^{-\alpha} .
\end{equation}
The asymptotic value of the scaling exponent $\alpha$ is obtained from the
analysis of the effective exponent $\alpha_{e}(N,\delta) \equiv
-\ln(P_\infty(mN,\delta)/P_\infty(N,\delta))/\ln m$ 
with constant $m=4$. 
If $P_\infty$ follows the power law, then the effective exponent
will converge to the scaling exponent in the $N\rightarrow\infty$ limit. 
Otherwise, it will converge to a trivial
value of $0$ or $1$. 

\begin{figure}[t]
\includegraphics*[width=\columnwidth]{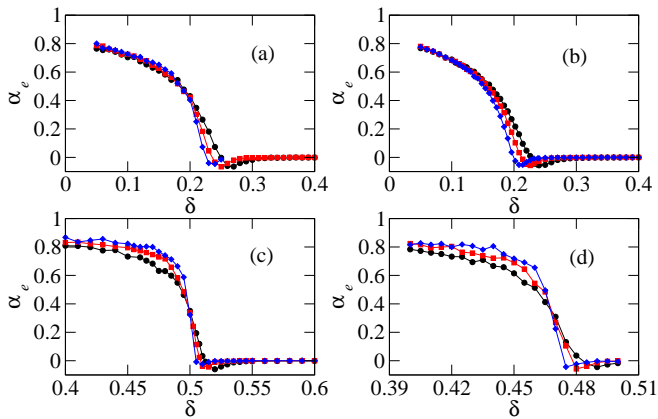}
\caption{Plot of the effective exponent $\alpha_e(N,\delta)$ 
in the correlated static model with $\mu=0.0$~(a) and $\mu=0.2$~(b), 
and in the static model with $\mu=0.0$~(c) and $\mu=0.2$~(d). Circle,
square, and diamond symbols are used for $N=(1,4,16)\times 10^5$, 
respectively.}
\label{fig4}
\end{figure}
In Figure~\ref{fig4}, we present a plot of $\alpha_e(N,\delta)$ 
for the static model and the correlated static model.
We find a significant difference
in finite-size-scaling behaviors of $\alpha_e$. 
First, consider the correlated static model shown in 
Figure~\ref{fig4}(a) and (b). 
At large values of $\delta$, $\alpha_e$ vanishes as $N$ increases, which 
indicates that the system is in the percolating phase with finite
$P_\infty$.
On the other hand, at small values of $\delta$, 
it converges to a nontrivial value $\alpha<1$
that varies continuously with $\delta$. 
This suggests that the correlated static
model is critical not only at the percolation threshold but also 
in the entire non-percolating phase. We estimate the percolation threshold
as the point at which the power-law scaling sets in. The results are 
$\delta_c = 0.19(1)$ at $\mu=0.0$ and $\delta_c = 0.15(1)$ at $\mu=0.2$. At
the transition point, the scaling exponent takes the value 
\begin{equation}
\alpha_c = 0.5(1)
\end{equation}
in both cases with $\mu=0.0$ and $0.2$.

The unusual scaling behavior of $P_\infty$ in the correlated static model
becomes clear when one compares it with that in the static model. 
Figure~\ref{fig4}(c) and (d) show that the static-model networks are 
critical only at the percolation thresholds, $\delta_c\simeq 0.50$ at 
$\mu=0.0$ and $\delta_c \simeq 0.47$ at $\mu=0.2$. 
The scaling exponent at the critical
point is given by $\alpha_c \simeq 0.33$. These numerical results are in
good agreement with the analytic results~\cite{Lee04}.

\begin{figure}[t]
\includegraphics*[width=\columnwidth]{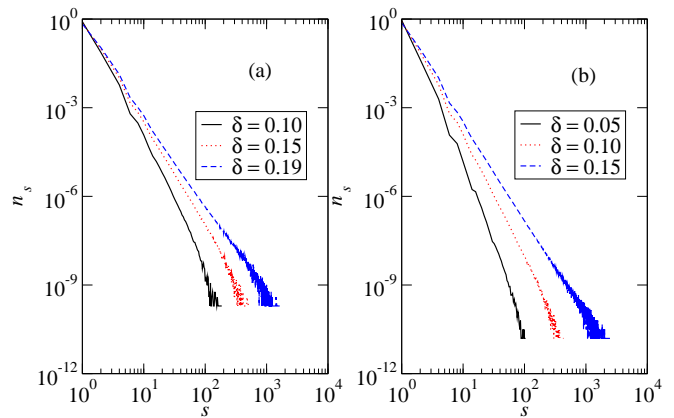}
\caption{Cluster size distribution, $n_s$, in the non-percolating phase of the correlated static model with $\mu=0.0$~(a) and $0.2$~(b). }
\label{fig5}
\end{figure}

In order to investigate further the criticality in the non-percolating phase, 
we study the cluster size distribution, $n_s$. Figure~\ref{fig5} shows 
$n_s$ at $\delta\leq \delta_c$ in the correlated static model with 
$\mu=0.0$ and $0.2$. We find that the cluster size distribution follows a
power-law distribution $n_s\sim s^{-\tau}$ with varying exponent $\tau$.
This result also suggests that the correlated static model is
critical in the entire non-percolating phase.

We also studied the percolation transition at larger values of 
$\mu$~(smaller value of $\lambda$). In the static
model, networks are always in the percolating phase~($\delta_c=0$) if
$\mu \geq 1/2$ or $\lambda \leq 3$~\cite{Lee04}. 
However, our numerical data show that the percolation threshold in the
correlated static model already vanishes at $\mu=1/4$~($\lambda=5$).
This implies that the correlated static model network is more robust 
than the static model network.

\section{Summary}\label{sec:4}
We have introduced a correlated static model 
and investigated the nature of the percolation transition.
Interestingly, the correlated static model is critical 
in the entire non-percolating phase below the percolation threshold. That
is, the percolation order parameter scales algebraically as $P_\infty\sim
N^{-\alpha}$ and the cluster size
distribution follows the power law as $n_s \sim s^{-\tau}$ with varying
exponents $\alpha$ and $\tau$. Such a type of percolation transition is 
also observed in the growing-network models and the assortative exponential 
random-graph model.
These results support the claim that the assortative DD correlation 
may give rise to a structural correlation which is responsible for novel 
type of percolation transition.

\begin{acknowledgments}
This work was supported by Korea Research Foundation Grant
funded by the Korean Government (MOEHRD, Basic Research Promotion Fund)
(KRF-2006-003-C00122). This work was also supported by KOSEF through the grant
No. R17-2007-073-01001-0.
\end{acknowledgments}

\end{document}